\documentclass{revtex4}
\usepackage{amsmath,amsthm}
\usepackage{setspace}
\usepackage{graphicx}
\newcommand{\bra}[1]{\langle#1|}
\newcommand{\ket}[1]{|#1\rangle}
\usepackage{amsfonts}

\theoremstyle{remark}

\theoremstyle{definition}

\theoremstyle{example}

\theoremstyle{notation}

\usepackage{amssymb}

\begin{document}
\draft
\title{Interpolation between phase space quantities with bifractional displacement operators}
\author{S. Agyo, C. Lei, A. Vourdas\\
Department of Computing,\\
University of Bradford, \\
Bradford BD7 1DP, United Kingdom}

\begin{abstract}
Bifractional displacement operators, are introduced by performing two fractional Fourier transforms on displacement operators.
They are shown to be special cases of elements of the group $G$, that contains both displacements and squeezing transformations.
Acting with them on the vacuum we get various classes of coherent states, which we call bifractional coherent states.
They are special classes of squeezed states which can be used for interpolation between various quantities in phase space methods. 
Using them we introduce bifractional Wigner functions $A\left (\alpha, \beta;\theta _{\alpha},\theta _{\beta}\right )$,
which are a two-dimensional continuum of functions, and reduce to Wigner and Weyl functions in special cases.
We also introduce bifractional $Q$-functions, and bifractional $P$-functions.
The physical meaning of these quantities is discussed.
\end{abstract}

\maketitle

\section{Introduction}

Coherent states have been studied extensively in the literature for a long time\cite {C1,C2,C3}.
They play a central role in phase space methods in quantum mechanics\cite{Z1,Z2}.
Various generalizations of coherent states have also been studied, especially in connection with groups like $SU(2)$, $SU(1,1)$, etc\cite{ii}. 

Two important operators in phase space methods, are the displacement operators and the parity operators\cite{G1,G2,V1,V2}.
They are related to each other through a two-dimensional Fourier transform. In this paper we replace the two Fourier transforms with two fractional Fourier transforms
\cite{F1,F2,F3,F4}, 
and we get new unitary operators $U(\alpha, \beta; \theta _{\alpha}, \theta _{\beta})$ which we call bifractional displacement operators.
Both displacement operators and parity operators, are special cases of these more general operators.
We show that $U(\alpha, \beta; \theta _{\alpha}, \theta _{\beta})$ 
are elements of the group $G=HW\rtimes SU(1,1)$ that contains both displacements and squeezing transformations.
The latter have been studied extensively in the literature, and in this paper we show that the operators
$U(\alpha, \beta; \theta _{\alpha}, \theta _{\beta})$, which are introduced with an interpolation motivation,
are elements of the group $G$ (but the general element of $G$ cannot always be written in the form
$U(\alpha, \beta; \theta _{\alpha}, \theta _{\beta})$).

Acting with the bifractional displacement operators on the vacuum, we get various classes of generalized coherent states (one class for each pair 
$(\theta _{\alpha}, \theta _{\beta})$), which we call bifractional coherent states. They are squeezed states, and here we 
study them briefly as a topic in its own right, because of their role in interpolations of different quantities in phase space methods.
Using the bifractional displacement operators, we also introduce
bifractional Wigner functions. Both the Wigner and Weyl functions are special cases of these more general functions.
Examples of such functions are calculated numerically.
Functions which interpolate between various phase space quantities, provide a deeper insight to current work on the phase space formalism.

In section 2 we review briefly fractional Fourier transforms, in order to define the notation.
In section 3 we introduce the bifractional displacement operators and show that they are elements of the group $G$.
In section 4 we introduce the bifractional coherent states.
In section 5 we use the  bifractional displacement operators 
and the bifractional coherent states, to define bifractional Wigner functions and bifractional $Q$-functions.
We conclude in section 6 with a discussion of our results.

\section{Fractional Fourier Transform}

The Fractional Fourier transform  on a function $f(x)$ is given by
\begin{eqnarray}\label{2}
f(x; \theta)&=&{\mathfrak F}(\theta )[f(x)]=\int \Delta (x,y;\theta)f(y)dy\nonumber\\
\Delta (x,y;\theta)&=&\left [\frac {1+i\cot{\theta}}{2\pi}\right ]^{1/2}\nonumber\\
&\times&\exp \left [\frac{-i(x^2+y^2)\cot\theta}{2}+\frac{ixy}{ \sin \theta}\right ] 
\end{eqnarray}

Below we present some special cases of the $\Delta (x,y;\theta)$:
\begin{eqnarray}
&&\Delta (x,y;0)=\delta (x-y)\nonumber\\
&&\Delta \left (x,y;\frac{\pi}{2}\right )=\frac{\exp (ixy)}{(2\pi)^{1/2}}\nonumber\\
&&\Delta (x,y;\pi)=\delta (x+y)
\end{eqnarray}
In the case $\theta=\pi/2$ the fractional Fourier transform reduces to the Fourier transform.
In the case $\theta=\pi$ it is the parity transform (later we introduce the parity operator $\Pi(0,0)$).
For later use we give the formula
\begin{eqnarray}\label{77}
\int dy\Delta (x,y;\theta _1)\Delta (y,z;\theta _2)=\Delta (x,z;\theta _1+\theta _2)
\end{eqnarray}

\section{Bifractional displacement operators}

We consider the usual position and momentum operators $\hat{x}, \hat{p}$ of the harmonic oscillator, and the displacement operators 
\begin{eqnarray}
D(\alpha, \beta) = \exp(i\sqrt{2}\beta\hat{x}-i\sqrt{2}\alpha\hat{p}).
\end{eqnarray}
The displaced parity operator is given by 
\begin{eqnarray}
\Pi(\alpha, \beta)&=&D\left (\frac{\alpha}{2},\frac{\beta}{2}\right )\Pi(0,0)D^\dagger \left (\frac{\alpha}{2},\frac{\beta}{2}\right )\nonumber\\
&=&D(\alpha, \beta)\Pi(0,0);\nonumber\\ \Pi(0,0)&=&\int dx\ket{x}\bra{-x}.
\end{eqnarray}
It is related with the displacement operator through the two-dimensional Fourier transform (e.g., \cite{G1,G2,V1,V2}) 
\begin{alignat}{2}\label{7}
&&\Pi(\alpha, \beta)=&\frac{1}{2\pi} \int  D(\alpha ', \beta ')\exp\left [i(\beta \alpha '-\beta '\alpha)\right ] d\alpha ' d\beta '\nonumber\\
&&=& \int d\alpha 'd\beta '\Delta \left (\beta, \alpha ';\frac{\pi}{2}\right )
\Delta \left (\alpha,-\beta ';\frac{\pi}{2}\right ) D(\alpha ', \beta ')
\end{alignat}

A generalization of the displaced parity operator is the following operator which we call bifractional displacement operator. The bifractional displacement operator is a unitary operator and defined as
\begin{alignat}{2}\label{8}
&&U(\alpha, \beta; \theta _{\alpha}, \theta _{\beta})=&|\cos (\theta_{\alpha}-\theta _{\beta})|^{1/2}
\int d\alpha 'd\beta '\Delta \left (\beta,\alpha ';\theta _{\beta}\right )\nonumber\\
&&&\times \Delta \left (\alpha,-\beta ';\theta _{\alpha}\right )
D(\alpha ', \beta ')\nonumber\\
&&[U(\alpha, \beta; \theta _{\alpha}, \theta _{\beta})]^{\dagger}=&U(-\alpha, -\beta; -\theta _{\alpha}, -\theta _{\beta})
\end{alignat}

Here we replaced the two Fourier transforms in Eq.(\ref{7}) with two fractional Fourier transforms.
We note that the two fractional Fourier transforms, use the variables $\alpha', \beta '$ which are related to position 
and momentum and are dual to each other. 
In this sense {\bf our  two-dimensional fractional Fourier transform
is not a straightforward generalization of a
one-dimensional fractional Fourier transform}.

The prefactor $|\cos (\theta_{\alpha}-\theta _{\beta})|^{1/2}$ in Eq.(\ref{8}) is important for unitarity.
The fact that this prefactor cannot be factorized as a function of $\theta _\alpha$ times
a function of $\theta _\beta$, is related to the fact that the variables $\alpha', \beta '$ are dual quantum variables.
Also for $\theta_{\alpha}-\theta _{\beta}=\pi/2$, the prefactor is zero and the integral diverges, and in numerical work below,
we do not go near this point.

The following are special cases:
\begin{eqnarray}
&&U\left (\alpha, \beta;0,0\right )=D(\beta, -\alpha)\nonumber\\
&&U\left (\alpha, \beta; \frac{\pi}{2}, \frac{\pi}{2}\right )=\Pi(\alpha, \beta)\nonumber\\
&&U\left (\alpha, \beta; \pi, \pi\right )=D(-\beta, \alpha)
\end{eqnarray}

\subsection{$U(\alpha,\beta; \theta _{\alpha}, \theta _{\beta})$ as special elements of the group $G$ of squeezing and displacement transformations}\label{FF}

The bifractional displacement operator is also given by
\begin{alignat}{1}\label{bto}
&U(\alpha,\beta; \theta _{\alpha}, \theta _{\beta})=
\exp (i\phi)\exp \left [i\tau(\hat{p}-\tan \theta _{\alpha}\hat {x}+\sigma)^2\right ]\nonumber\\
&\hspace{1cm}\times \exp \left (i\frac{\hat{x}^2}{\cot \theta _{\alpha}}-i\frac{\sqrt{2}\alpha \hat{x}}{\cos \theta_{\alpha}}
\right)\nonumber\\
&\tau=\frac{\cos\theta _{\alpha}\sin \theta _{\beta}}{\cos( \theta _{\alpha}-\theta _{\beta})};\;\;\;\;
\sigma=\frac{\alpha}{\sqrt{2}\cos \theta _{\alpha}}-\frac{\beta}{\sqrt{2}\sin \theta_{\beta}}
\nonumber\\&\phi=-\frac{1}{2}(\theta _{\alpha}+\theta _{\beta})-\frac{1}{2}(\alpha ^2\cot \theta _{\alpha}+\beta ^2\cot \theta_{\beta})+\frac{\alpha ^2}{\sin 2\theta _\alpha}.
\end{alignat}

The proof is complex and is based on the integration of Eq.(\ref{8}). In the integration we are careful with the ordering of the operators.

We recall here that the operators
$\hat {x}^2$,
$\hat {p}^2$,
$\hat {x}\hat {p}+\hat {p}\hat {x}$,
$\hat {x}$,
$\hat {p}$,
${\bf 1}$, 
form a closed structure under commutation. Therefore the
\begin{alignat}{1}
&T(a_1,a_2,a_3,a_4,a_5,a_6)=\nonumber\\
&\exp[a_1 \hat {x}^2+a_2\hat {p}^2+a_3(\hat {x}\hat {p}+\hat {p}\hat {x})+a_4\hat {x}+a_5\hat {p}+a_6{\bf 1}]
\end{alignat}
form a group G which is the semidirect product of the Heisenberg Weyl group $HW$ of displacements, by the
$SU(1,1)$ group of squeezing transformations: $G=HW\rtimes SU(1,1)$.
The operators $U(\alpha,\beta; \theta _{\alpha}, \theta _{\beta})$ depend on four parameters, and they are
special cases of the operators $T(a_1,a_2,a_3,a_4,a_5,a_6)$. 
But clearly the general element $T(a_1,a_2,a_3,a_4,a_5,a_6)$ which depends on six parameters cannot always be written as
$U(\alpha,\beta; \theta _{\alpha}, \theta _{\beta})$ which depends on four parameters.

\section{Bifractional coherent states}

Given a pair $(\theta _{\alpha}, \theta _{\beta})$ we introduce the following set of `bifractional coherent states':
\begin{eqnarray}
{\cal C}(\theta _{\alpha}, \theta _{\beta})=\{\ket{\alpha, \beta; \theta _{\alpha}, \theta _{\beta}}=U(\alpha, \beta; \theta _{\alpha}, \theta _{\beta})\ket{0}\}
\end{eqnarray}
In the special cases that $\theta _{\alpha}=\theta _{\beta}=0$ we get 
\begin{eqnarray}
\ket{\alpha, \beta; 0,0}=U(\alpha, \beta; 0,0)\ket{0}=D(\beta, -\alpha)\ket{0}.
\end{eqnarray}
In the special cases that $\theta _{\alpha}=\theta _{\beta}=\frac{\pi}{2}$ we get 
\begin{eqnarray}
\ket{\alpha, \beta;\frac{\pi}{2},\frac{\pi}{2}}=U(\alpha, \beta; \frac{\pi}{2},\frac{\pi}{2})\ket{0}=D(\alpha , \beta)\ket{0}.
\end{eqnarray}
Therefore in these special cases we get the standard (Glauber) coherent states, and
\begin{eqnarray}
\ket{-\beta, \alpha; 0,0}=\ket{\alpha, \beta ;\frac{\pi}{2}, \frac{\pi}{2}}=\ket{\alpha, \beta }
\end{eqnarray}

Since the bifractional operator is unitary, the coherent states in the set ${\cal C}(\theta _{\alpha}, \theta _{\beta})$
are Glauber coherent states with respect to the operators 
\begin{eqnarray}
b(\theta _{\alpha}, \theta _{\beta})&=&U(0,0; \theta _{\alpha}, \theta _{\beta})a[U(0,0; \theta _{\alpha}, \theta _{\beta})]^{\dagger}\nonumber\\
b^\dagger (\theta _{\alpha}, \theta _{\beta})&=&U(0,0; \theta _{\alpha}, \theta _{\beta})a^\dagger [U(0,0; \theta _{\alpha}, \theta _{\beta})]^{\dagger}.
\end{eqnarray}
But they have novel non-trivial properties with respect to $a, a^{\dagger}$.

The coherent states in the set ${\cal C}(\theta _{\alpha}, \theta _{\beta})$
satisfy the resolution of the identity
\begin{eqnarray}\label{res}
\frac{1}{2\pi}\int d\alpha d\beta \ket{\alpha, \beta; \theta _{\alpha}, \theta _{\beta}}\bra {\alpha, \beta; \theta _{\alpha}, \theta _{\beta}}={\bf 1}
\end{eqnarray}
From this follows that an arbitrary state $\ket{g}$ can be written as
\begin{alignat}{1}
&\ket{g}=\int d\alpha d\beta\ket{\alpha, \beta; \theta _{\alpha}, \theta _{\beta}}g(\alpha, \beta; \theta _{\alpha}, \theta _{\beta});\nonumber\\
&g(\alpha, \beta; \theta _{\alpha}, \theta _{\beta})=\frac{1}{2\pi}\bra{\alpha, \beta; \theta _{\alpha}, \theta _{\beta}}g\rangle 
\end{alignat}

We note that the 
\begin{alignat}{1}
&{\cal B}(z)=g(\alpha, \beta;\frac{\pi}{2},\frac{\pi}{2})\exp [\frac{1}{2}(\alpha ^2+\beta ^2)];\nonumber\\
&z=\alpha +i\beta
\end{alignat}
is a Bargmann function with respect to the Glauber coherent states $\ket {\alpha, \beta ; \frac{\pi}{2}, \frac{\pi}{2}}$.

We next consider the special case $\theta _{\beta}=0$ and
calculate the Bargmann functions ${\cal B}(z; \theta _{\alpha}, 0)$ \cite{B1,B2,V2} for the coherent states in the set ${\cal C}(\theta _{\alpha}, 0)$. 

\begin{eqnarray}\label{67}
&&{\cal B}(z; \theta _{\alpha},0)=|\cos \theta_{\alpha}|^{1/2}\exp(Az^2+Bz+\Gamma)\nonumber\\
&&A=-\frac{1}{2(1+i\cot \theta_{\alpha})}\nonumber\\
&&B=\beta+\frac{\alpha}{\sin \theta _{\alpha}(1+i\cot \theta_{\alpha})}\nonumber\\
&&\Gamma=-\frac{1}{2}(\beta ^2+\alpha ^2)
\end{eqnarray}
This result should be compared and contrasted with the Bargmann function for the squeezed states
\begin{alignat}{1}
&\ket{w;r,\theta}=\exp\left[-\frac{1}{4}re^{-i\phi}(a^{\dagger})^2+\frac{1}{4}re^{i\phi}a^2\right]
\exp(wa^{\dagger}-w^*a)\ket{0}\nonumber\\
&a=2^{-1/2}(\hat {x}+i \hat {p});\;\;\;\;\;a^{\dagger}=2^{-1/2}(\hat {x}-i \hat {p})
\end{alignat}
which is\cite{V2}
\begin{eqnarray}\label{676}
&&{\mathfrak B}(z)=(1-|a|^2)^{1/4}\exp \left(\frac{1}{2}az^2+bz+c\right )\nonumber\\
&&a=-\tanh \left(\frac{r}{2}\right )e^{-i\phi}\nonumber\\
&&b=w(1-|a|^2)^{1/2}\nonumber\\
&&c=-\frac{1}{2}a^*w^2-\frac{1}{2}|w|^2.
\end{eqnarray}
If in Eq.(\ref{676}) we replace $a$ with $2A$, $b$ with $B$, and $c$ with $\Gamma$, we get Eq.(\ref{67}).

The coherent states in the set ${\cal C}(\theta _{\alpha}+\phi _{\alpha}, \theta _{\beta}+\phi _{\beta})$ are related to the coherent states in the set
${\cal C}(\theta _{\alpha}, \theta _{\beta})$ 
through the fractional Fourier transform
\begin{alignat}{1}
&\ket{\alpha, \beta; \theta _{\alpha}+\phi _{\alpha}, \theta _{\beta}+\phi _{\beta}}
=\frac{|\cos (\theta_{\alpha}+\phi _{\alpha}-\theta _{\beta}-\phi _{\beta})|^{1/2}}{|\cos (\theta_{\alpha}-\theta _{\beta})|^{1/2}}\nonumber\\&\times
\int d\alpha 'd\beta '\Delta \left (\beta, \beta ';\phi _{\beta}\right )
\Delta \left (\alpha , \alpha ';\phi _{\alpha}\right )
\ket{\alpha ', \beta '; \theta _{\alpha}, \theta _{\beta}}.
\end{alignat}

\subsection{Statistical properties of the coherent states $\ket{\alpha, \beta; \theta _{\alpha}, 0}$}

\begin{figure}
\caption{The uncertainty $\sigma _{pp}$, the $g^{(2)}$ and the average number of photons $<n>$ 
as a function of $\theta _{\alpha}$ (in rads), for the coherent states  $\ket{2,2; \theta _{\alpha}, 0}$}
\includegraphics[scale = 0.3]{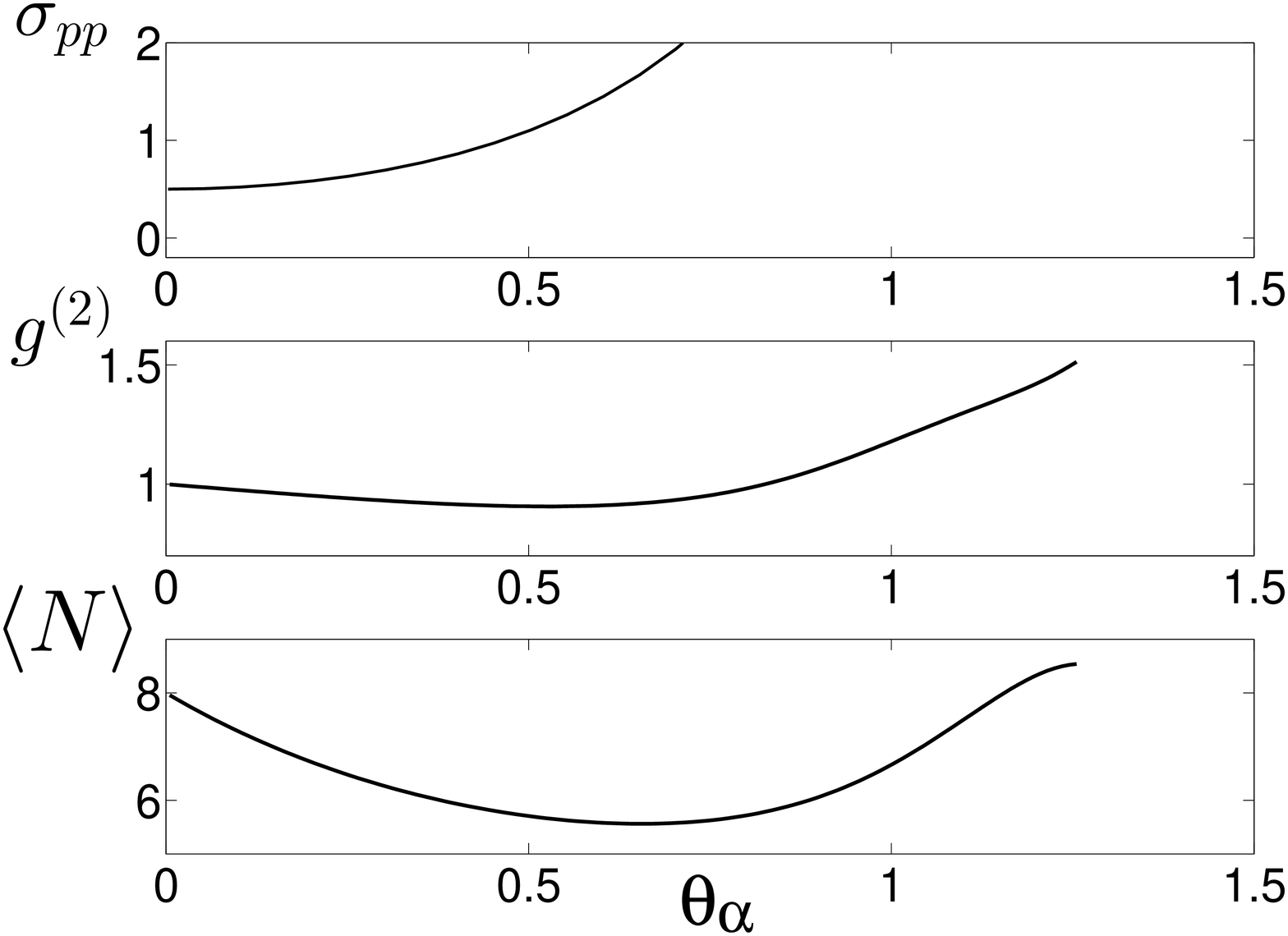}
\label{f1}
\end{figure}

For the coherent states $\ket{\alpha, \beta; \theta _{\alpha}, 0}$ (with $\theta _{\beta}=0$) we have calculated
the wavefunction $f(x)$ using the relation
\begin{alignat}{1}
&f(x)=\pi ^{-3/4}\exp \left (-\frac{x^2}{2}\right )\int dp\;{\cal B}((x+ip)\sqrt{2}; \theta _{\alpha},0)\exp(-p^2)
\end{alignat}
and the Bargmann function of Eq.(\ref{67}). We found
\begin{alignat}{1}
&f(x)=|\cos \theta _\alpha|^{1/2}\pi ^{-1/4}\left (\frac{1}{1+2A}\right )^{1/2}\exp \left (\frac{\kappa x^2+2^{3/2}Bx +\lambda}{2+4A}\right )\nonumber\\
&\kappa=2A-1;\;\;\;\;\;\lambda=2\Gamma +4A\Gamma -B^2.
\end{alignat}
We then calculated the uncertainties
\begin{alignat}{1}
&\sigma_{xx}=<x^2>-<x>^2;\;\;\;\;\;<x^n>=\int dx x^n|f(x)|^2\nonumber\\
&\sigma_{pp}=<p^2>-<p>^2;\;\;\;\;\;<p^n>=\int dx[f(x)]^*(-i\partial _x)^nf(x)\nonumber\\
&\sigma_{xp}=<\frac{1}{2}(xp+px)>-<x><p>;\nonumber\\
&<\frac{1}{2}(xp+px)>=-\frac{i}{2}+\int dx[f(x)]^*x(-i\partial _x)f(x)
\end{alignat}
and similarly for $\Delta p$. As expected for squeezed states,
we found numerically that for all angles $\theta _{\alpha}$, the Robertson-Schr\"{o}dinger relation gives
\begin{eqnarray}
\sigma_{xx}\sigma_{pp}-\sigma_{xp}^2=\frac{1}{4}.
\end{eqnarray}
We also found analytically that
\begin{eqnarray}
<x>=\beta \sqrt {2};\;\;\;\;\;<x^2>=2\beta ^2+\frac{1}{2};\;\;\;\;\sigma_{xx}=\frac{1}{2}.
\end{eqnarray}
It is seen that these quantities do not depend on $\alpha, \theta _{\alpha}$ (in the case $\theta _\beta=0$ considered in this section).
The $\sigma_{pp}$ has been calculated numerically for the case $\alpha=\beta =2$. 
In Fig.\ref{f1} we plot the $\sigma_{pp}$ as a function of $\theta _{\alpha}$.
Also in this example
\begin{eqnarray}
\sigma_{xp}^2=\frac{1}{2}\sigma_{pp}-\frac{1}{4}
\end{eqnarray}

Using a Taylor expansion we have expressed the Bargmann function ${\cal B}(z; \theta _{\alpha},0)$ of Eq.(\ref{67}), with $\alpha=\beta =2$, as
\begin{eqnarray}
{\cal B}(z; \theta _{\alpha},0)&=&\sum _{n=0}^\infty \frac {a_nz^n}{\sqrt{n!}}
\end{eqnarray}
Numerically we have truncated the series at $30$. We have found that in this case $\sum |a_n|^2=0.999$.
We have then calculated the quantities
\begin{eqnarray}
<n^\nu>=\sum _{n=0}^{30} n^\nu |a_n|^2;\;\;\;\;g^{(2)}=\frac{<n^2>-<n>}{<n>^2}
\end{eqnarray}
In Fig.\ref{f1} we show $<n>$ and $g^{(2)}$ as a function of $\theta _\alpha$.
It is seen that for $\theta _\alpha<0.8$ we have antibunching ($g^{(2)}<1$).

\section{Interpolations between phase space quantities}
In the first part of this section, we introduce a two-dimensional continuum of functions $A\left (\alpha, \beta;\theta _{\alpha},\theta _{\beta}\right )$, 
which we call bifractional Wigner functions, and which have the Weyl and Wigner functions, as special cases. In the second  part, we introduce 
bifractional $Q$-functions, and bifractional $P$-functions.

\subsection{Bifractional Wigner functions}

For a trace class operator $\Theta$, the Weyl function is given by
\begin{eqnarray}
\widetilde {W}(\alpha , \beta |\Theta) &=& {\rm Tr}[\Theta D(\alpha , \beta )]
\end{eqnarray}
and the Wigner function is given by
\begin{eqnarray}
W(\alpha , \beta |\Theta) &=& {\rm Tr}[\Theta \Pi(\alpha , \beta )]
\end{eqnarray}
From Eq.(\ref{7}) it follows that 
the Wigner and Weyl functions are related through the two-dimensional Fourier transform (e.g.,\cite{W1,W2}):
\begin{alignat}{1}\label{78}
{W}(\alpha, \beta|\Theta) = \frac{1}{2\pi} \int {\widetilde W}(\alpha ',\beta '|\Theta)\exp\left [i(\beta \alpha '-\beta '\alpha)\right ] d\alpha ' d\beta '
\end{alignat}
Using Eq.(\ref{8}) we define the bifractional Wigner function
\begin{alignat}{2}\label{wi}
&&A(\alpha, \beta;\theta_{\alpha},\theta_{\beta}|\Theta) =& {\rm Tr}(\Theta  U(\alpha, \beta; \theta _{\alpha}, \theta _{\beta})]\nonumber\\
&&=&
|\cos (\theta_{\alpha}-\theta _{\beta})|^{1/2}\int 
\Delta \left (\beta , \alpha ';\theta _{\beta}\right )\nonumber\\
&&&\times \Delta \left (\alpha , -\beta ';\theta _{\alpha}\right )  
 {\widetilde W}(\alpha ', \beta '|\Theta)d\alpha 'd\beta '
\end{alignat}
If it is obvious which operator we use, we ommit $\Theta$ in the notation.
$A(\alpha, \beta;\theta_{\alpha},\theta_{\beta}|\Theta)$ includes both the Wigner and Weyl function as special cases:
\begin{eqnarray}
&&A\left (\alpha, \beta;0,0|\Theta\right )={\widetilde W}(\beta, -\alpha |\Theta)\nonumber\\
&&A\left (\alpha, \beta; \frac{\pi}{2}, \frac{\pi}{2}|\Theta \right )=W(\alpha, \beta |\Theta)\nonumber\\
&&A\left (\alpha, \beta; \pi, \pi|\Theta \right )={\widetilde W}(-\beta, \alpha |\Theta)
\end{eqnarray}
We note that \cite{V3} has considered a single fractional Fourier transform between Wigner and Weyl 
functions and this gives generalized Wigner functions that depend on one angle.
Here we use a double fractional Fourier transform and we get the bifractional Wigner functions that depend on two angles.
As explained earlier,  our  two-dimensional fractional Fourier transform
is not a straightforward generalization of a
one-dimensional fractional Fourier transform.

\subsection{Bifractional $Q$-functions and bifractional $P$-functions}

$q$-functions (or Husimi functions) for a trace class operator $\Theta$, are defined as
\begin{eqnarray}
q(\alpha, \beta|\Theta)
=\bra{\alpha, \beta}\Theta\ket{\alpha, \beta}
\end{eqnarray}
We define bifractional $Q$-functions with respect to the bifractional coherent states as follows:
\begin{eqnarray}
Q(\alpha, \beta; \theta _{\alpha}, \theta _{\beta}|\Theta)
=\bra{\alpha, \beta; \theta _{\alpha}, \theta _{\beta}}\Theta\ket{\alpha, \beta; \theta _{\alpha}, \theta _{\beta}}
\end{eqnarray}
Using the resolution of the identity in Eq.(\ref{res}) we get
\begin{eqnarray}
\frac{1}{2\pi}\int d\alpha d\beta\;Q(\alpha, \beta; \theta _{\alpha}, \theta _{\beta}|\Theta)
={\rm Tr}\Theta.
\end{eqnarray}
Clearly
\begin{eqnarray}
&&Q(\alpha, \beta;0,0|\Theta)=q(\beta, -\alpha|\Theta);\nonumber\\
&&Q(\alpha, \beta; \frac{\pi}{2}, \frac{\pi}{2}\Theta)=q(\alpha, \beta|\Theta)
\end{eqnarray}
The bifractional $Q$-functions are generalizations of the $q$-functions.

We also introduce the bifractional $P$-function $P(\alpha, \beta; \theta _{\alpha}, \theta _{\beta}|\Theta)$ 
with respect to the bifractional coherent states, as
\begin{alignat}{1}
&\Theta =\frac{1}{\pi}\int d\alpha\;d\beta P({\alpha, \beta; \theta _{\alpha}, \theta _{\beta}}|\Theta)\ket{\alpha, \beta; \theta _{\alpha}, \theta _{\beta}}\bra{\alpha, \beta; \theta _{\alpha}, \theta _{\beta}}
\end{alignat}
In the case $\theta _{\alpha}=\theta _{\beta}=\frac{\pi}{2}$ we get the ordinary $P$-function $P(\alpha, \beta)$.
The $P({\alpha, \beta; \theta _{\alpha}, \theta _{\beta}}|\Theta)$ is given by
\begin{alignat}{1}
&P({\alpha, \beta; \theta , \phi}|\Theta)=\nonumber\\
&\frac{1}{\pi}\exp (|\alpha |^2+|\beta |^2)\int d\gamma d \delta 
\exp [2i(\beta \gamma-\alpha \delta)]\exp (|\gamma |^2+|\delta |^2)\nonumber\\
&\hspace{1cm}\times
\bra{-\gamma, -\delta; \theta , \phi}\Theta\ket{\gamma, \delta; \theta , \phi}
\end{alignat}
The proof of this is the same as in \cite{M}. 
This is because the overlap $\langle{\gamma, \delta; \theta _{\alpha}, \theta _{\beta}}\ket{\alpha, \beta; \theta _{\alpha}, \theta _{\beta}}$
does not depend on $\theta _{\alpha}, \theta _{\beta}$.

\section{Discussion}
The parity and displacement operators are important tools in phase space methods.
The parity operators are a two-dimensional Fourier transform of the displacement operators (Eq.(\ref{7})).
We have replaced the two Fourier transforms, with two fractional Fourier transforms,
and we got the bifractional displacement operators $U(\alpha, \beta; \theta _{\alpha}, \theta _{\beta})$.
We stressed the importance for unitarity, of the prefactor $|\cos (\theta_{\alpha}-\theta _{\beta})|^{1/2}$ in these operators.
We have shown that the $U(\alpha, \beta; \theta _{\alpha}, \theta _{\beta})$ are special cases of the squeezing operators
acting on displacement operators,
which are used here for interpolation purposes.

Acting with the bifractional displacement operators on the vacuum, we get 
the bifractional coherent states, which are special cases of squeezed states. 
We have studied the uncertainties and statistical properties of these states.
They can be used to define generalized phase space quantities.
For example we have used them to define generalizations of the $Q$-functions.

Using the bifractional displacement operators we introduced the bifractional Wigner functions in Eq.(\ref{wi}).
Both the Wigner and Weyl functions are special cases of this more general function.

The work provides a deeper insight into the phase space methods.

\newpage

\end{document}